\documentclass[aps,amssymb,amsmath,singlecolumn,showpacs,showkeys]{revtex4}
\usepackage{graphicx}
\newcommand{\qref}[1]{(\ref{#1})}

\newcommand{\ket}[1]{\left|{#1}\right\rangle}

\def\sn{{\rm sn\,}}
\def\cn{{\rm cn\,}}
\def\dn{{\rm dn\,}}
\include{graphics}
\begin{document}


\title{{\bf Designing bound states in a band as a model for a quantum network}}
\author{
S. Sree Ranjani$^{a}$\footnote{ranjani@physics.iitm.ernet.in},
A. K. Kapoor$^{b}$\footnote{akksp@uohyd.ernet.in},
 and P. K. Panigrahi$^{c}$\footnote{prasanta@prl.ernet.in}}
\affiliation{$^a$ Department of Physics, Indian Institute of Technology
  Madras, Chennai, 600 036, India\\$^{b}$School of Physics, University
  of Hyderabad,   Hyderabad 500 046, India \\
$^c$ Physical Research Laboratory Navrangpura, Ahmedabad, 380 009,
India}

\begin{abstract}
We provide a model of a one dimensional quantum network, in the
framework of a lattice using  Von Neumann and Wigner's idea of bound
states in a continuum. The localized states acting as qubits are
created by a controlled deformation of a periodic potential. These
wave functions lie at the band edges and are defects in a lattice. We
propose that these defect states, with atoms trapped in them, can be
realized in an optical lattice and can act as a model for a quantum network.
\end{abstract}
\pacs{03.65Ge, 03.67Lx}
\keywords{Lam\'e equation, band structure, bound states, qubits,
  quantum networks.}
\maketitle

\section{Introduction}
   In classical computation, bits  $0$ and $1$ are used to represent
data. Logic gates are employed for computation  and data manipulation. At
present, considerable work is being done to develop quantum
computers. It is expected that they will increase the speed and
efficiency of computation [\ref{nel}] compared to a classical
computer. The  basic building blocks in quantum computing are known as
quantum bits or qubits and are represented by the eigenstates of a system. For
example, qubits are described  by the up and down states of spin-half
particles, represented by $\ket{0}$ and $\ket{1}$. The
unitary operators play the role of logic gates and we can perform
various operations similar to classical computing. The major areas of
interest in the field of quantum computation  are the development of
models, having states which can be used as qubits and the construction
of unitary operators to manipulate these qubits. It is important that
these states should be well isolated from the outside environment to
avoid environmental decoherence.

    In this paper, we propose a theoretical model for preparing
qubits {\it i.e.}, $|0>$ and $|1>$, using periodic potentials. These
potentials are characterized by a band spectrum,
which contains bands of allowed energies interspersed  with forbidden
energy gaps. The wave functions are not integrable and they extend to spatial
infinity. Hence, they cannot be envisaged as qubits.
We will explicitly demonstrate that a periodic potential can be
appropriately deformed to accommodate bound states in its band
spectrum.  This deformation can be treated as the perturbation of
the original potential. The effect of which is the
creation of the localized states which can be perceived as
defects in the lattice. Atoms trapped in these defects can be used to
describe qubits. This treatment of the trapped atoms in optical
lattices, as qubits in a quantum network, is similar to the
quantum computation model proposed by Angelakis {\it
et. al.,}[\ref{dimi}]. Here, they have used photonic crystals to
confine photons in the defect states created inside the band gap, to
represent qubits. In quantum networking  atoms are trapped in the
nodes and are used to store information. We propose that an
array of such localized states in the deformed optical lattice, with
atoms trapped in them, can be used as a quantum network [ \ref{cirac},
  \ref{asoka}].

At present, various optical lattices are routinely realized in the
laboratories and atoms can be trapped in a potential well with
relative ease. For our model, we consider one of the well - known
family of periodic potentials namely, the Lam\'e potentials
[\ref{mag}, \ref{ars}],
\begin{equation}
V(x) = j(j+1)m\,\sn^2(x,m).   \label{ll1}
\end{equation}
The function  $\sn(x,m)$ is the Jacobi elliptic function [\ref{han},
  \ref{witt}] with elliptic modulus $0 <m <1$. These potentials are
exactly solvable for integer values of $j$ and for a given
$j$ there are $(2j+1)$ band-edges. Both the form of these band-edge
solutions and explicit solutions for smaller values of $j$ are  given
in [\ref{ars}, \ref{ppqhj}, \ref{khlam}].  The Lam\'e potential has
been proposed as a model for quasi 1 - d confinement of  Bose - Einstein
condensates (BEC) in a standing light wave [\ref{bron}]. Possible
application of BECs for quantum computation
are currently being explored and the bound states created by deforming
a potential like the Lam\'e potential may have useful implications.

    The method used to construct these bound states is the same as that
used by Pappademos {\it et. al.,} to construct quantum mechanical
bound states in a classical continuous energy spectrum. Such states
were first discovered by Von Neumann and Wigner
[\ref{von}] and many such examples were found later [\ref{stil} -
\ref{anz}]. It became clear that such states arose due to the
delicate interplay between the rate at which the oscillatory potential
falls off and the time taken by the various maxima of the potential to
interact and create a bound state.  Existence of such states was
reported by Capasso {\it et. al.,} in semiconductor hetrostructures
[\ref{cap}].

    SUSYQM [\ref{wit},
\ref{khajp}]  was applied to generate  bound states in the
continuum of the energy  spectrum for a spherically symmetric
potential [\ref{pap}, \ref{pkp}]. We apply the same technique to periodic
potentials and construct bound states in their band spectrum, with
the same energies as the band-edge energies. The
solutions thus obtained are located at the band-edges and not inside the
forbidden energy gap as discussed in [\ref{dimi}, \ref{dav}, \ref{frank}].

  In the next section, we give an overview of SUSYQM  and briefly describe
the steps involved in constructing a bound state in the continuum (BIC). In
section III, we use the same procedure to construct bound states for the Lam\'e
potential. In the last section, we present our conclusions.\\

\section{ Supersymmetric quantum mechanics}

       For a given 1 - d potential $V(x)$ with eigenfunctions $u_n(x)$ and
eigenvalues $E_n$ $(n=0,1,2...)$,  we can generate a new family of
potentials $\tilde{V}(x;\lambda)$, isospectral to $V(x)$. The parameter
$\lambda$ is used to label the potentials in the isospectral family and takes
values lying in the range $\lambda >0$ and $\lambda <-1$. Setting
$\hbar = 2m =1$ we can write the superpotential as,
\begin{equation}
W(x) = -\frac{u^{\prime}_0}{u_0}.     \label{e1}
\end{equation}
The original potential $V(x)$ can be expressed in terms of $W(x)$ as,
\begin{equation}
V(x) = W^2(x)-W^{\prime}(x).     \label{e2}
\end{equation}
Its isospectral partner  $\tilde{V}(x;\lambda)$ is given by
\begin{equation}
\tilde{V}(x;\lambda) = W^2(x)+W^{\prime}(x).                      \label{e3}
\end{equation}
Let $\tilde{u}_n(x)$  be the eigenfunctions of $\tilde{V}(x;\lambda)$
$(n=1,2...)$. We have $\tilde{u}_n(x) = Au_n(x)$ where $A = d/dx
+ W(x)$. (Hence,  $A^{\dag}=-d/dx + W(x)$). Note that
$\tilde{u}_0(x)$, the ground state cannot be obtained in this manner
since $Au_0(x) =0 $. Thus, $\tilde{V}(x;\lambda)$ is isospectral to
$V(x)$, except that its spectrum does not contain
$\tilde{u}_0(x)$. Hence, to introduce the ground state into its
spectrum and form a complete set of eigenstates, we need to find the
most general superpotential  $\tilde{W}(x)$, so that
\begin{equation}
\tilde{V}(x;\lambda) = \tilde{W}^2(x)+\tilde{W}^{\prime}(x). \label{e4}
\end{equation}
We can now show that
\begin{equation}
\tilde{W}(x) = W(x) +\frac{d}{dx}\ln(I_0(x)+\lambda) \label{e5}
\end{equation}
where
\begin{equation}
I_0= \int_{0}^{x}u^{2}_0(y)dy.           \label{e6}
\end{equation}
Thus, in this process of reinstating the ground state, we obtain the
expression for the potential $\tilde{V}(x;\lambda)$  in terms of $I_0$ as,
\begin{equation}
\tilde{V}(x;\lambda) =V(x)-2[\ln(I_0+\lambda)]^{\prime\prime}
\,\,=\,\,V(x)-\frac{4u_0u^{\prime}_0}{I_0+\lambda}
+\frac{2u^{4}_0}{(I_0+\lambda)^2}.    \label{e7}
\end{equation}
 Note that $\tilde{V}(x;\lambda)$ is $V(x)$ plus a perturbative
term and $\lambda$ can be take as the perturbative parameter. The
perturbed potential is isospectral to the old potential  and its
eigenstates are given below in Eqs. \qref{e8} and \qref{e9}.

    In standard SUSYQM, isospectral families of potentials
which allow only bound states have been constructed, and $u_0(x)$ was taken
to be the ground state.  In [\ref{pkp}], the above method was
generalized to the case where the potentials have a continuous energy
spectrum and $u_0(x)$ has been taken to be any non-singular eigenstate of
$V(x)$. This procedure was used to construct BIC for the spherically
symmetric potential and the results were
stated as a theorem. We refer the reader to [\ref{pap}] and the
references therein for further details and merely reproduce the
result here. \\

{\it Let $u_0(x)$ and $u_1(x)$ be any two nonsingular solutions
  of the Schr\"odinger equation for the potential V(x) corresponding to
  arbitrarily selected energies $E_0$ and $E_1$ respectively. Construct a new
  potential  $\tilde{V}(x;\lambda)$ as prescribed by
  Eq. \qref{e7}. Then, the two
  functions
\begin{equation}
\tilde{u}_0(x) = \frac{u_0}{I_0+\lambda}    \label{e8}
\end{equation}
and
\begin{equation}
\tilde{u}_1(x) = (E_1 - E_0)u_1(x) + \tilde{u}_0(x)W_r(u_0(x),u_1(x))
\label{e9}
\end{equation}
are solutions of the Schr\"odinger equation for the new potential
$\tilde{V}(x;\lambda)$, corresponding to the same energies $E_0$ and
$E_1$. Here $W_r(u_0(x),u_1(x))$ is the Wronskian.}

   Note that the original potential $V(x)$  had no integrable solutions
but the new potential has one square-integrable solution
$\tilde{u}_0(x)$,  with the rest being non -  integrable. The
creation of the bound state can be
elucidated by the fact that $I_0$ in Eq. \qref{e6} diverges owing to the
non-integrability of $u_0$. Hence, as  $I_0 \rightarrow \infty$,
$\tilde{u}_0(x) \rightarrow 0$, resulting in a square-integrable
wave function in the continuum.

   We can create another bound state by using the non-normalizable
state  $\tilde{u}_1(x)$ in place of $u_0(x)$ and deforming
$\tilde{V}(x;\lambda)$, using  the same procedure described
above. We then obtain a potential
$\tilde{\tilde{V}}(x;\lambda,\lambda_1)$, isospectral to $V(x)$ which has
two bound states in the continuous spectrum with energies $E_0$ and
$E_1$. The parameter $\lambda_1$ is a real number lying in the range
$\lambda_1 >0$ and $\lambda_1 <-1$.

    The expressions for the new potential $\tilde{\tilde{V}}(x)$ and the two
square-integrable states $\tilde{\tilde{u}}_0(x)$ and
$\tilde{\tilde{u}}_1(x)$,  with energies $E_0$ and $E_1$ respectively
are given by
\begin{equation}
\tilde{\tilde{V}}(x)= \tilde{V}(x)-2[\ln(I_1+\lambda_1)]^{\prime\prime}
\,\,=\,\,\tilde{V}(x)-\frac{4\tilde{u}_1\tilde{u}^{\prime}_1}{I_1+\lambda_1}
+\frac{2\tilde{u}^{4}_1}{(I_1+\lambda_1)^2},\label{e10}
\end{equation}

\begin{equation}
\tilde{\tilde{u}}_0(x)= (E_0 - E_1)\tilde{u}_0(x) +
\tilde{\tilde{u}}_1(x)W_r(\tilde{u}_1(x),\tilde{u}_0(x)) \label{e11}
\end{equation}
and
\begin{equation}
\tilde{\tilde{u}}_1(x)= \frac{\tilde{u}_1}{I_1+\lambda_1}  \label{e12}
\end{equation}
where
\begin{equation}
I_1= \int_{0}^{x}\tilde{u}^{2}_1(y)dy  \label{e13}
\end{equation}
and $W_r(\tilde{u}_1(x),\tilde{u}_0(x))$ is the Wronskian. In the next
section, we apply the above technique to periodic potentials
on the half-line and construct bound states in the band spectrum. For this
purpose, we use the band-edge wave functions to deform the original
periodic potential. The bound states thus created have the
same  band-edge energies.

\section{The Lam\'e Potential}

      The Lam\'e potential (Eq. \qref{ll1}) with $j=2$ is,
\begin{equation}
V(x) = 6m\sn^{2}(x,m).  \label{l1}
\end{equation}
It has two bands  and a continuum. The expressions for the five band
- edge  wave functions and energies are given below [\ref{ars}] [
\ref{ppqhj}, \ref{khlam}] [\ref{footnote}] with $\psi_0(x) $ and
$\psi_1(x)$  representing the lower and upper band - edge wave
functions of the first band and so on. We have
\begin{eqnarray}
\psi_0(x)  = 3m +3 -\delta - 3m \sn^{2}(x,m) \,, \,\, E_0 = 2\delta-2m -2,
 \label{l2}\\
 \psi_1(x) = \cn (x,m)\, \dn (x,m) \,, \,\, E_1 = m+1,  \label{l3}\\
\psi_2(x) = \dn (x,m) \,\sn (x,m) \, , \,\, E_2 = 4m+1,  \label{l4}\\
\psi_3(x) = \cn (x,m)\,\sn (x,m)  \, , \,\, E_3 = m+4,    \label{l5}\\
\psi_4(x) = 3m +3 -3\delta - 3m \sn^2(x,m) \, ,\,\, E_4 = 2\delta + 2m +2,
\label{l6}
\end{eqnarray}
where, the functions $\cn(x,m)$ and $\dn(x,m)$ are the Jacobi elliptic
functions with modulus parameter $m$. For constructing the bound
states, we examine only the half-line problem and hence consider the
band-edge wave functions  which vanish at the origin. In the above
given expressions, only Eqs. \qref{l4} and \qref{l5} which represent
the lower and upper band edge wave functions of the second band,
satisfy this condition, since $\sn(0,m) = 0$.
   We follow the steps described in the previous section and construct
two bound states with energies $E_2$ and $E_3$. The entire procedure is done
numerically and we give the plots of the deformed potential and bound states
thus obtained, in the sequel.

\begin{figure}[h]
\includegraphics[width=6cm]{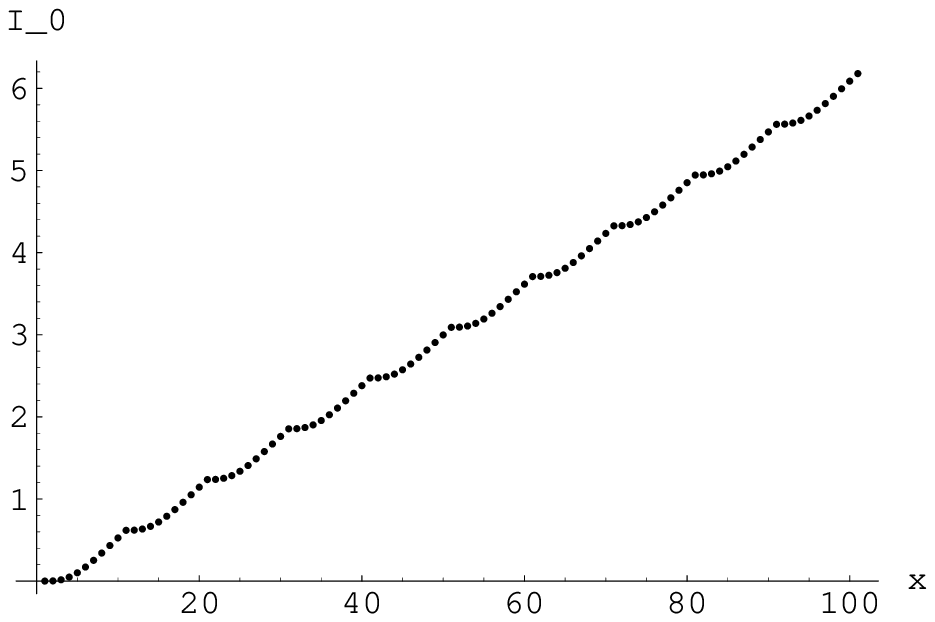}
\caption{Plot of the diverging integral $I_0$ versus $x$. }
\label{I0}
\includegraphics[width=6cm]{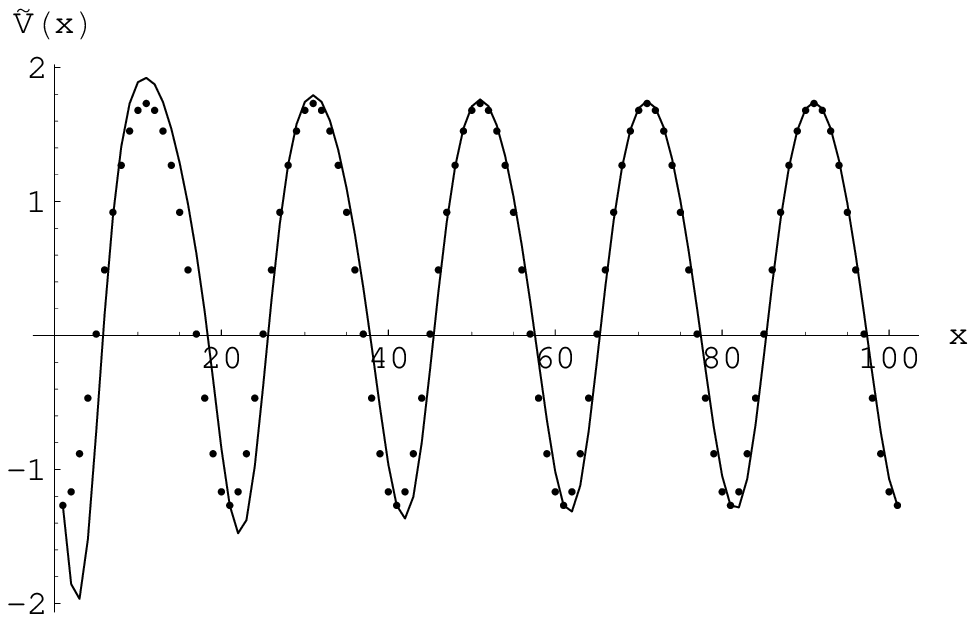}
\caption{Deformed potential $\tilde{V}(x)$  versus $x$, with $\lambda
  = 1$. The dotted line represents the original potential $V(x)$. }
\label{vthil}
\includegraphics[width=6cm]{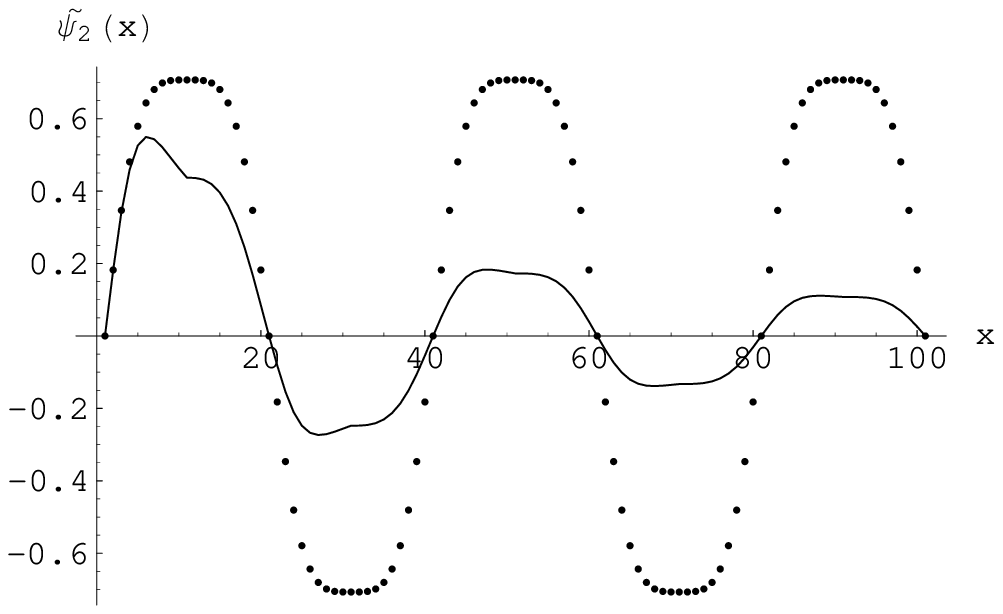}
\caption{Deformed wave function $\tilde{\psi}_2(x)$  versus $x$, with
  $\lambda = 1$. The dotted line represents the original band-edge
  wave function $\psi_2(x)$. } \label{cap2}
\label{u0thil}
\includegraphics[width=6cm]{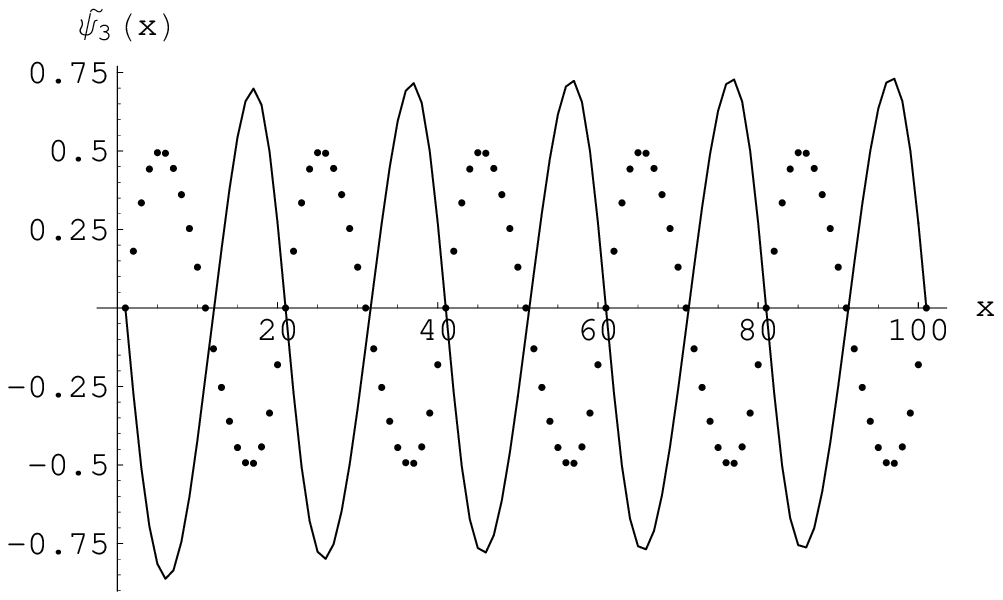}
\caption{Deformed wave function $\tilde{\psi}_3(x)$ versus $x$, with $\lambda =
1$. The dotted line represents the original band-edge wave function
$\psi_3(x)$. }
\label{u1thil}
\end{figure}

      We deform the potential given in Eq. \qref{l1}, using the
expression for  $\psi_2(x)$ in
Eq. \qref{l4}, to obtain a one-parameter bound state solution which
depends on the parameter $\lambda$ . For this purpose, we first plot
$I_0$ versus $x$ in  Fig. \ref{I0}. As expected $I_0$ turns out to be a
diverging integral. Using $I_0$ and Eqs. \qref{e7}, \qref{e8} and
\qref{e9}, we
plot the deformed potential $\tilde{V}(x)$ and the deformed wave
functions $ \tilde{\psi}_2(x)$ and $\tilde{\psi}_3(x)$. These are give
in Figs. \ref{vthil}, \ref{u0thil} and \ref{u1thil} respectively. For
comparison,  the original potential and wave functions  are plotted
in dotted line.

\begin{figure}[h]
\includegraphics[width=6cm]{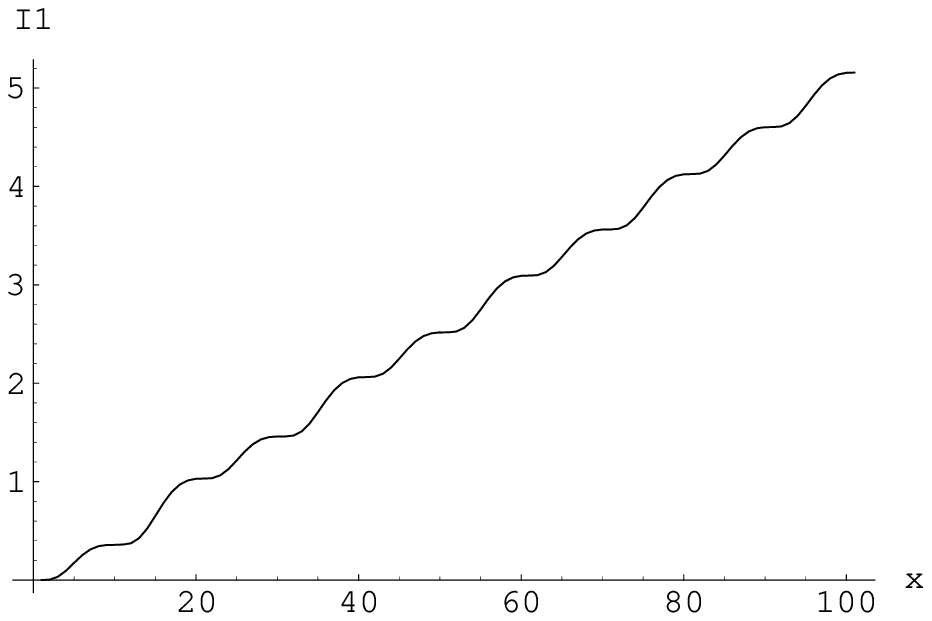}
\caption{Plot of the diverging integral $I_1$  versus $x$. }
\label{I1}
\includegraphics[width=6cm]{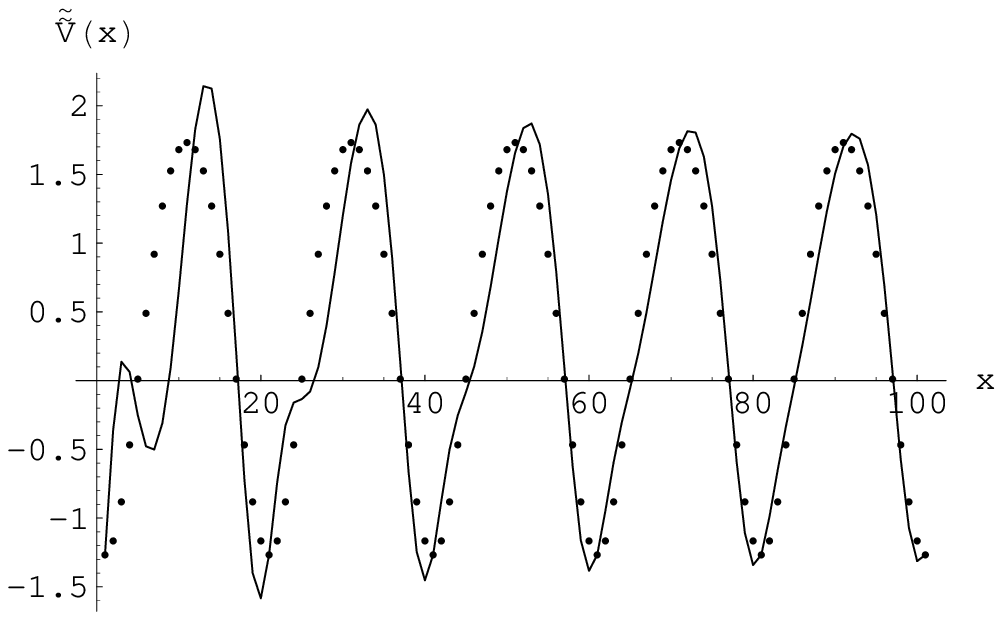}
\caption{Deformed potential $\tilde{\tilde{V}}(x)$  versus $x$, with
  $\lambda_1 = 1$. The dotted line represents the original potential $V(x)$. }
\label{v2thil}
\includegraphics[width=6cm]{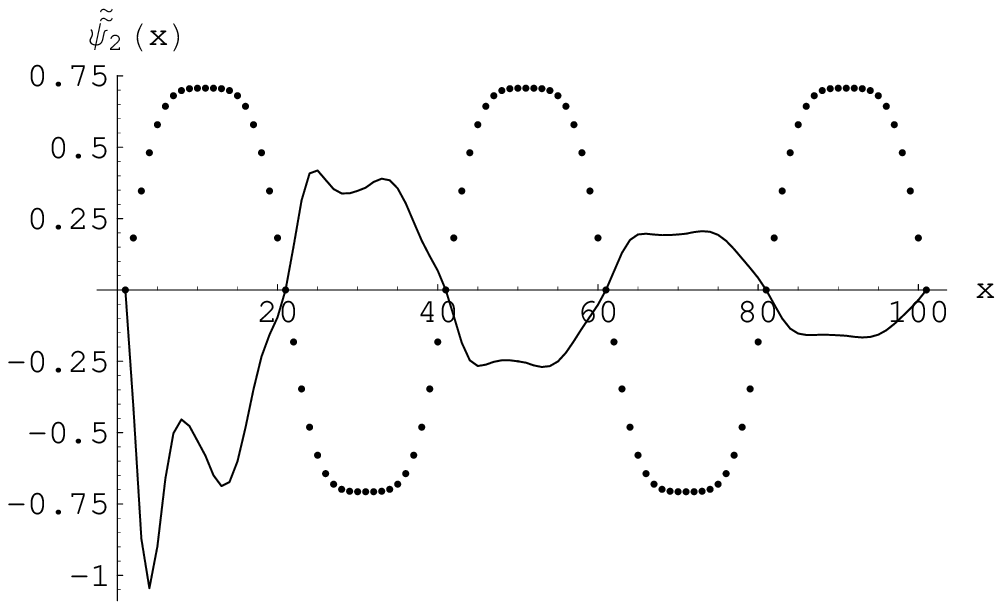}
\caption{Deformed wave function $\tilde{\tilde{\psi}}_2(x)$  versus
  $x$, with $\lambda_1 = 1$. The dotted line represents the original
  band-edge wave function $\psi_2(x)$. }
\label{u02thil}
\includegraphics[width=6cm]{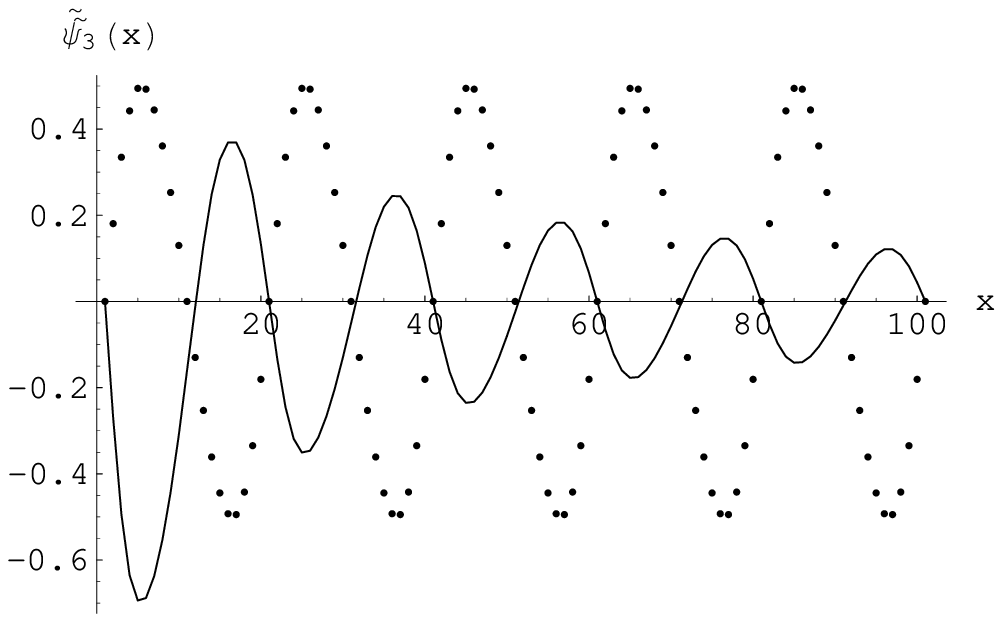}
\caption{Deformed wave function $\tilde{\tilde{\psi}}_3(x)$  versus
  $x$, with $\lambda _1= 1$. The dotted line represents the original
  band-edge wave function
$\psi_3(x)$. }
\label{u12thil}
\end{figure}

   It is clear from Figs. \ref{u0thil} and \ref{u1thil} that
$\tilde{\psi}_2(x)$ is a normalizable state and $\tilde{\psi}_3(x)$
is not normalizable. Thus, with this deformation we have obtained
only one bound state. In order to construct two bound states, we
deform $\tilde{V}(x)$  with $\tilde{\psi}_3(x)$ using Eqs \qref{e10}
- \qref{e13}. Plots of $I_1$,  $\tilde{\tilde{V}}(x)$,
$\tilde{\tilde{\psi}}_2(x)$ and $\tilde{\tilde{\psi}}_3(x)$ versus
$x$ are given in Figs. \ref{I1}, \ref{v2thil}, \ref{u02thil} and
\ref{u12thil} respectively.

\begin{figure}[h]
\includegraphics[width=10cm]{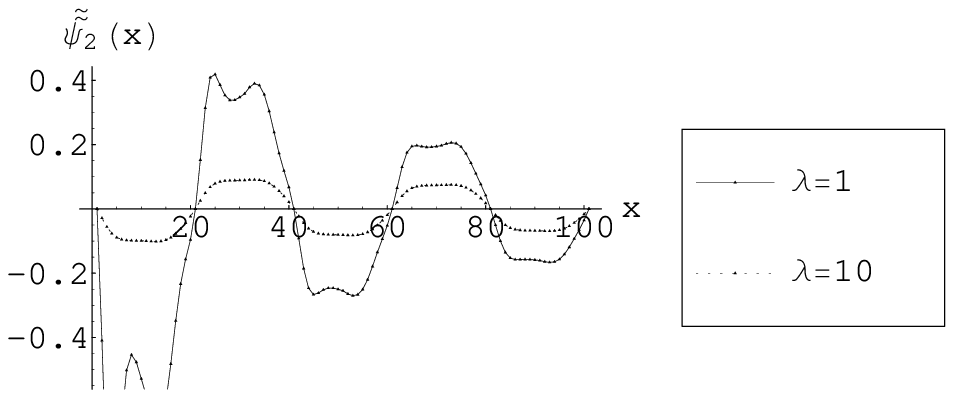}
\caption{Deformed wave function $\tilde{\tilde{\psi}}_2(x)$ for $\lambda_1 =
\lambda =1$  (dotted line) and for $\lambda_1 =\lambda =10$ (thick line).}
\label{lam2}

\includegraphics[width=10cm]{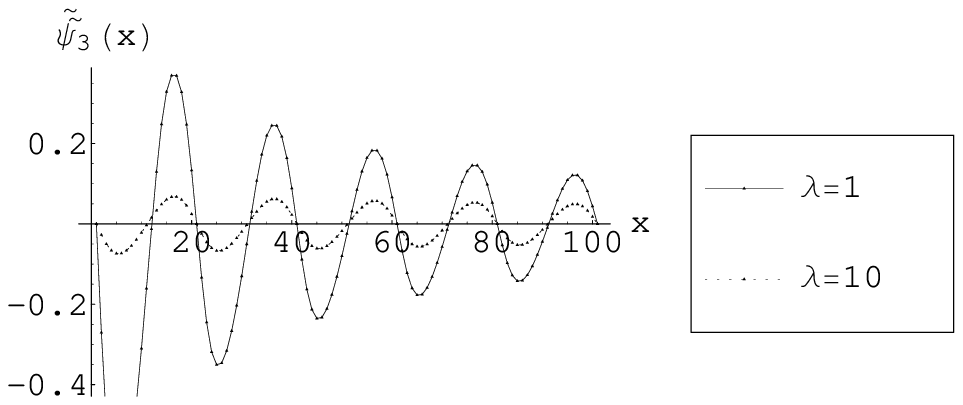}
\caption{Deformed wave function $\tilde{\tilde{\psi}}_3(x)$ for $\lambda _1=
  \lambda = 1$ (dotted line) and for $\lambda_1= \lambda =10$ (thick line).}
\label{lam1}
\end{figure}

From Figs.\ref{u02thil} and \ref{u12thil}, it is clear that both
$\tilde{\tilde{\psi}}_2(x)$ and $\tilde{\tilde{\psi}}_3(x)$ are
integrable. These states have energies $E_2$ and $E_3$ and the
potential $\tilde{\tilde{V}}(x)$ is isospectral to the original
potential $V(x)$. The deformed potential and the bound states depend
on the parameters $\lambda$ and $\lambda_1$ and as we increases
their values, the deformed states and the potential tend towards the
corresponding original states and the potential. In Figs. \ref{lam2}
and \ref{lam1}, we give  the plots of the deformed wave functions
$\tilde{\tilde{\psi}}_2(x)$ and $\tilde{\tilde{\psi}}_3(x)$, for two
different values of $\lambda$ and $\lambda_1$. (Without loss of
generality  we have set $\lambda = \lambda_1 = 1$ and $\lambda
=\lambda_1= 10$ in these figures).

\section{Conclusions}

    We have shown that we can construct bound
states in the band spectrum of a periodic potential using SUSYQM. It is clear
from the above procedure that we can create a class of  bound states  by
successively deforming the potential $V(x)$, provided there
exist band-edge wave functions, {of the original potential}, which
satisfy the boundary condition $\psi_n(0) = 0$.

   We can use these square-integrable states as qubits in
quantum computation. The localized states with atoms trapped in them
can be treated as the qubits $\ket{0}$ and $\ket{1}$.
If it is possible to get $n$ such defect states in the optical lattice
we have an array of qubits which can be used as an optical network
[\ref{asoka}]. The perturbative parameters can be used to adjust the
overlap of these trapped atoms and also to control the deformation of
the periodic potential. Moreover, these localized states are in the
band and hence, protected from the external influences. Since optical
lattices are easy to create and manipulate in the laboratory, this can
be a useful model for quantum networking [\ref{palmer}, \ref{jak},
  \ref{shloma}].

   In conclusion, we have shown that it is possible to deform a
periodic potential to accommodate localized states at the band
edges. It is proposed that by trapping atoms in these states we can
construct qubits and an array of $n$ such qubits can be used as a
quantum network.

{\bf Acknowledgments :} S. S. R. thanks  S. Lakshmi Bala for
useful comments and acknowledges support from the Department of Science
and Technology, India, under project No. SP/S2/K-14/2000. \\


{\bf References}

\begin{enumerate}

\item {\label{nel}}  M. A. Nielsen and I. L. Chuang, {\it Quantum
  Computation and Quantum Information} (Cambridge Univ. Press, 2002).

\item {\label{dimi}} D. G. Angelakis, M. F. Santos, V. Yannopappas and
  A. Ekert, quanth - ph/0410189.

\item {\label{cirac}} J. I. Cirac and P. Zoller, {\it Phys. Rev. A.}
  {\bf 74}, 4091 (1995).

\item {\label{asoka}} A. Biswas and G. S. Agarwal. {\it Phys. Rev. A.}
  {\bf 70}, 02232 (2004).

\item {\label{mag}} W. Magnus and S. Wrinkler, {\it Hills Equation}
  (Interscience Publishers, New York, 1966).

\item {\label{ars}} F. M. Arscot, {\it Periodic Differential Equations}
  (Pergamon, Oxford, 1964).

\item {\label{han}} H. Hancock, {\it Theory of Elliptic Functions} (Dover
Publications, Inc, New York, 1958).

\item {\label{witt}} E. Whittaker and G. N. Watson, {\it A Course of Modern
  Analysis} (Cambridge Univ. Press, Cambridge, 1963).

\item {\label{ppqhj}} S. Sree Ranjani, A. K. Kapoor and P. K. Panigrahi,
  Mod. Phys. Lett. A {\bf 19}, No. {\bf 27}, 2047 (2004); quant - ph/0312041

\item {\label{khlam}} A. Khare and U. Sukhatme, J. Math, Phys. {\bf 40}, 5473
  (1999).

\item {\label{dun}} G. V. Dunne and J. Feinberg, Phys. Rev. D {\bf 57}, 1271
  (1998).

\item {\label{bron}} J. C. Bronski, L. D. Carr, B. Deconinck and
  J. N. Kutz, Phys. Rev. Lett. {\bf 86}, 1402 (2001).

\item {\label{von}} J. Von Neumann and E. Wigner, Phys. Z. {\bf 30}, 465
  (1929).

\item {\label{stil}} F. H. Stillinger and D. R. Herrick,  Phys. Rev. A {\bf
  11}, 446 (1975).

\item {\label{mey}} N. Meyer - Vernet, Am. J. Phys. {\bf 50}354 (1982).

\item {\label{anz}} A. Khelashvili and N. Kiknadze, J. Phys. A: Math. Gen. {\bf
    29}, 3209 (1996).

\item {\label{cap}} F. Capasso, C. Sirtori, J. Faist, D. L. Sivco, Sung -  Nee,
  G. Chu and A. Y. Cho, Nature, {\bf 358}, 565 (1992).

\item {\label{wit}} E. Witten, Nucl. Phys. B {\bf 185}, 513 (1981).

\item {\label{khajp}} R. Dutt, A. Khare and U. Sukhatme, Am. J. Phys. {\bf 56},
 163 (1988).

\item {\label{pap}} J. Pappademos, U. Sukhatme and A. Pagnamenta,
  quant - ph/9305336.

\item {\label{pkp}} P. K Panigrahi and U. Sukhatme, Phys. Lett. A
  {\bf 178}, 251 (1993).

\item {\label{dav}} D. J. Fern\'andez C, B. Mielknik, O. Rosas - Ortiz
  and B. F. Samsonov,  J. Phys. A: Math. Gen. {\bf
    35}, 4279 (2002).

\item {\label{frank}} F. Szmulowicz, Am. J. Phys. {\bf 72} (11) 1392 (2004).

\item {\label{palmer}} R. N. Palmer, C. Moura Alves and D. Jaksch,
  quanth - ph/0506059.

\item {\label{jak}} D. Jaksch, H. -J. Briegel, J. I. Cirac,
  C. W. gardiner, and P. Zoller, {\it Phys. Rev. Lett.} {\bf 82}, 1975
  (1999).

\item {\label{shloma}} S. E. Sklarz, I. Friedler, D. J. Tanner,
  Y. B. Band and C. J. Williams, {\it Phys. Rev. A.} {\bf 66}, 053620 (2002)

\item{\label{footnote}} The solutions given in
these references [\ref{ars}, \ref{ppqhj}, \ref{khlam}]are for
supersymmetric potential, where a constant is added to the potential
to make the ground state energy zero.

\end{enumerate}

\end{document}